\documentclass[11pt,a4paper]{article}
\pdfoutput=1
\usepackage{jheppub}
\usepackage{graphicx}
\usepackage{amssymb}
\usepackage{amsmath}
\usepackage{topcapt}
\usepackage{epstopdf}
\DeclareMathOperator*{\jj}{\it j}

\preprint{DIAS-STP-14-09\\}

\title{Commuting Quantum Matrix Models}

\author[a]{Veselin G. Filev}
\author{and}
\author[b]{Denjoe O'Connor}

\affiliation[a,b]{School of Theoretical Physics, Dublin Institute for Advanced Studies\\
10 Burlington Road, Dublin 4, Ireland.}
\emailAdd{denjoe@stp.dias.ie}
\emailAdd{vfilev@stp.dias.ie}

\abstract{We study a quantum system of $p$ commuting matrices and find that such a quantum system {\it requires} an explicit curvature dependent potential in its Lagrangian for the system to have a finite energy ground state. In contrast it is possible to avoid such curvature dependence in the Hamiltonian. We study the eigenvalue distribution for such systems in the large matrix size limit. A critical r\^ole is played by $p=4$. For $p\ge4$ the competition between eigenvalue repulsion and the attractive potential forces the eigenvalues to form a sharp spherical shell.}


\subheader{}

\begin{document}
\maketitle

\section{Introduction}
Matrix quantum mechanics began \cite{Brezin:1977sv} with the problem
of counting planar diagrams in field theory. In ref.~\cite{Brezin:1977sv} a
single matrix was quantised and shown to be equivalent to a free Fermi
gas. In the large $N$ (matrix size) limit the eigenvalue density
was obtained and found to have similar features to pure matrix models.
The eigenvalues are confined to the finite domains such that
$E_F-V(\lambda)$ is positive and the eigenvalue density is given by
$\rho(\lambda)=\sqrt{2 E_F-2 V(\lambda)}$ in these domains with the
Fermi energy $E_F$ determined from the normalisation of the eigenvalue
density. For a quartic potential the model undergoes a
one-cut two-cut phase transition \cite{Shimamune:1981qf} when the quadratic 
term is made sufficiently negative and other features are 
broadly in line with the pure matrix model.

Interest in matrix quantum mechanics blossomed with the advent of
the AdS/CFT correspondence and multi-matrix models are
fundamental to current understanding of M-theory.

Although a non-perturbative formulation of M-theory in terms of its
fundamental degrees of freedom is still lacking, the best candidate
for such a formulation appears to be the infinite matrix size limit of
a matrix model of some kind. The leading candidate for such a
formulation is the BFFS model \cite{Banks:1996vh,Townsend:1995kk}
which was conjectured to capture the entire dynamics of M-theory.
Relatives of this model such as the BMN model \cite{Berenstein:2003gb}
or models derived from the ABJM model
\cite{Aharony:2008gk,Kovacs:2013una} are also considered possible
viable candidates for such a non-perturbative formulation.  All of
these conjectured formulations of M-theory are regularised versions of
the supermembrane and are matrix quantum mechanical systems. They are
based on the matrix regularisation of membranes introduced by Hoppe
\cite{Hoppe:PhDThesis1982} and extended to the supermembrane in
\cite{Townsend:1995kk} and \cite{de Wit:1988ig}.  They also arise as
dimensionally reduced 4-dimensional or 3-dimensional supersymmetric
field theories. 

Many of these models will have regimes where commuting matrices play a
r\^ole. It has been suggested by Berenstein \cite{Berenstein:2005aa}
that, in order to count $1/8$-BPS states, a matrix model of commuting
matrices is needed. 

Aside from the work of Berenstein \cite{Berenstein:2005aa} there is
little known about commutative quantum matrix models.  Our goal in this
article is to understand the quantisation and quantum mechanics of a
system of commuting matrices.  

Our first observation is that the configuration space of such matrix
models is curved.  The consequence of this is that there is no a
priori unique quantisation of the system. This is a classic problem
when the configuration space is curved and was clarified by B. De Witt
in his seminal article \cite{DeWitt:1957at}. There De Witt found that,
if one considers a classical dynamical system in a curved
configuration space, the corresponding quantum Hamiltonian has an
additional potential given by $\frac{\hbar^2}{6}{\cal R}$, where
${\cal R}$ is the scalar curvature of the configuration
space\footnote{For a more recent discussion see \cite{Cheng:1972yb}.}.
Other values for the coefficient of the scalar curvature have been
proposed \cite{DeWitt:1957at,Kleinert:1989rr} but since it is the
quantum system that is primary the coefficient can in principal have
any value. 

For the system of commuting matrices we therefore investigate the
general case and allow the coefficient have any value and comment on
the different possibilities. 

We begin our paper with a fresh derivation of the standard results of
matrix quantum mechanics. Using a WKB approximation for the single
particle wavefunctions we obtain a direct derivation of the eigenvalue
density.  We then establish the equivalence of the Bosonic and
Fermionic quantisations for the Gaussian potential. We are then in a
position to discuss the higher dimensional case where Bosonic
quantisation is the relevant one.

The principal results of this paper are:
\begin{itemize}
\item The configuration space of $p$-commuting Hermitian matrices with $p>1$
  is curved.  We compute the curvature of this space and establish
  that it has a curvature singularity when two eigenvalues
  approach. This singularity has a negative scalar curvature for
  $p>1$ independent of $N$.
\item If no explicit curvature dependence is included in the action for
  the system then a path integral quantisation of the system gives rise to
  a curvature term in the Hamiltonian. We find the resulting system
  has no finite energy ground state.
\item If the Hamiltonian is taken as primary and no explicit curvature
  dependence is included in this Hamiltonian then for Gaussian
  commuting matrices we find 
    \begin{enumerate}
     \item[(a)] For $p=2$ the eigenvalue distribution is uniform
       within a disc of unit radius.
     \item[(b)] For $p=3$ the eigenvalues are confined to a ball but
       have divergent density as the boundary is approached, though
       the distribution is integrable being
       $\rho_3(r)=\frac{3}{2\pi^2\sqrt{\frac{2}{3}-\vec{r}^2}}$.
     \item[(c)] For $p\ge4$ competition between 
eigenvalue repulsion and the attractive potential forces the eigenvalues
to form a shell with eigenvalue distribution $\rho_p(\vec
x)=\frac{2^{p/2}}{\Omega_{p-1}}\,\delta(1/2-\vec x^2)$ where $\Omega_{p-1}$ is
the volume of the unit $p-1$-sphere. Note the critical r\^ole played by the $p=4$ case. 
    \end{enumerate}
\item We find that if we quantize the $p$-commuting Hermitian matrices
  with $\frac{\xi}{6}{\cal R}$ as the scalar curvature contribution to
  the Hamiltonian then for $\xi \ge 1$ the system has no finite energy
  ground state for any polynomial potential. For $\xi <1 $ the
    system is stable and for Gaussian potentials the effect of the
    curvature term is to scale the radius by a factor of $({1-\xi})^{1/4}$. As $\xi\rightarrow 1$ we see the
    radius goes to zero and the distribution collapses.
\end{itemize}

As an aside we compute the scalar curvature of a generic squashed 
flag manifold where the squashing radii are given by the distances between 
eigenvalues. The scalar curvature of such a flag is always positive. It is 
the orthogonal complement of the flag in the configuration space of 
commuting matrices that gives the negative curvature. 
The space can be envisaged as a warped cone over the flag.

The layout of the paper is as follows: In section two we review the
one matrix case giving a novel and explicit derivation of the
eigenvalue density. We then establish, in the case of a quadratic
potential, the equivalence of the Fermionic and Bosonic
quantisations. The Bosonic quantisation will be the one relevant to
the subsequent sections. In section three we study commuting matrices
observing, to begin with, that the configuration space of such a
system is curved. We relegate the derivation of the curvature to
Appendix A which includes a derivation of the curvature of a generic
squashed flag manifold.  We then establish, for curvature potential
$\frac{\xi}{6}{\cal R}$, that when $\xi\ge 1$ the system has no finite
energy ground state.  In section 3.1 we discuss the range
$\xi<1$ and find that for a quadratic confining potential the
system is well defined and the eigenvalues are distributed uniformly
within a disc for $p=2$, while for $p=3$ they are confined to a
finite domain with the density increasing towards the boundary in a
divergent but integral form.  A special r\^ole is played by $p=4$ in
that it is the lowest dimension in which the eigenvalues are
concentrated in a spherical shell. For all $p>4$ the eigenvalues
form a spherical shell with the radius specified by
$R_p^2=\sqrt{1-\xi\,}/{2}$ and as $\xi$ approaches one the extent of the
eigenvalues goes to zero. 

We end the paper with discussion and
set the problem in the wider context by relating the system to
Calogero-Sutherland-Marchioro models and for $p=2$ to anyonic
systems.

\section{One matrix quantum model}

In this section we revisit the quantisation of a one matrix model
first presented in \cite{Brezin:1977sv}. Our goal is to perform a
semiclassical quantisation, in the fermionic description of
\cite{Brezin:1977sv}. For the special case of gaussian potential we
also demonstrate the equivalence of the bosonic quantisation of the
model. It is the bosonic quantisation which admits a natural
generalisation to the case with more than one commuting matrix.

Our starting point is the classical Lagrangian:
\begin{equation}\label{lag-1}
{\cal L} ={\rm tr}\left(\frac{1}{2}\dot X^2-V(X)\right)\ ,
\end{equation}
where $X$ is an $N\times N$ time dependent Hermitian matrix. The dot
represents a time derivate and V(X) is a potential which can be taken to
be of the form:
\begin{equation}
V(X)=\frac{1}{2}X^2+\frac{g}{N}X^4+\cdots\ .
\end{equation}
Hermitian matrices involve complex entries and the space of Hermitian
matrices is endowed with the flat metric:
\begin{equation}\label{metric-1}
ds^2={\rm tr} \left(dXdX^{\dagger}\right)={\rm tr}\left(dXdX\right)
\end{equation}
We could proceed and directly quantise the model by using
the Hamiltionian \cite{Brezin:1977sv}:
\begin{eqnarray}
H&=&-\frac{1}{2}\nabla^2+V\,\\
\nabla^2&=&\sum_i\frac{\partial^2}{\partial X_{ii}^2}+\frac{1}{2}\sum_{i < j}\frac{\partial^2}{\partial {\rm Re }\,X_{ij}^2}+\frac{\partial^2}{\partial {\rm Im }\,X_{ij}^2}\ .
\end{eqnarray}
But our interest will be in a system of commuting matrices, where the
elements of the matrices are not independent and in this case when the
constraints are solved we will see that commuting matrices have a curved
configuration space.  It is therefore most convenient to perform the
quantisation in the ``curved'' coordinates obtained by diagonalising
the matrices.  These complications are not necessary in the one matrix
model, where the configuration space is flat and diagonalisation is
just a change of coordinates. However, we will find it instructive to
follow a similar path in the quantisation of the one matrix model.

Diagonalising the matrix $X$ with a unitary
transformation $U$ so that $X=U\Lambda U^{\dagger}$ we arrive at the
following expression for the metric (\ref{metric-1}):
\begin{equation}\label{metric-2}
ds^2={\rm tr}\left(d\Lambda^2+ [\Lambda,\theta]^2\right)=\sum_id\lambda_i^2+2\sum_{i<j}(\lambda_i-\lambda_j)^2\theta_{ij}\bar{\theta}_{ij}\, 
\end{equation}
where $\theta =U^{\dagger}dU$ are the Maurer--Cartan left invariant
forms and $\lambda_i=\Lambda_{ii}$. Furthermore, in these coordinates
the Lagrangian (\ref{lag-1}) is given by:
\begin{equation}
{\cal L} =\frac{1}{2}{\rm tr}\left(\dot\Lambda^2+[\Lambda,\dot\theta]^2    \right)-{\rm tr}V(\Lambda)=\frac{1}{2}\sum_i\dot\lambda_i^2+\sum_{i<j}(\lambda_i-\lambda_j)^2\dot\theta_{ij}\bar{\dot\theta}_{ij}-\sum_iV(\lambda_i)\ ,
\end{equation}
where $\dot\theta =U^{\dagger}\dot U$. Note that the potential is independent on the $\theta$'s (and $U$) and they are cyclic variables. Our next step is to define generalised momenta:
\begin{equation}
\Pi_{\lambda_i} =\frac{\partial {\cal L}}{\partial \dot\lambda_i}=\dot\lambda_i\ ,~~\Pi_{\theta_{ij}} =\frac{\partial {\cal L}}{\partial\, \dot{\bar{\theta}}_{ij}}=(\lambda_i-\lambda_j)^2\dot\theta_{ij}\ ,~~\Pi_{\bar{\theta}_{ij}} =\frac{\partial {\cal L}}{\partial\, \dot{\theta}_{ij}}=\bar{\Pi}_{\theta_{ij}}\ , \text{for }i<j\ .
\end{equation}
For the corresponding Hamiltonian we obtain: 
\begin{equation}\label{hamil-01}
H=\frac{1}{2}\sum_i \Pi_{\lambda_i}^2+\sum_{i<j}\frac{\Pi_{\theta_{ij}}\Pi_{\bar{\theta}_{ij}}}{(\lambda_i-\lambda_j)^2}+\sum_iV(\lambda_i)=\frac{1}{2}g^{ab}\Pi_a\Pi_b+\sum_iV(\lambda_i)\ ,
\end{equation}
where $a,b$ are collective indices for the $\lambda_i, \theta_{ij}$
and $\bar\theta_{ij}$ components and $g^{ab}$ is the inverse of the
metric (\ref{metric-2}). The general prescription for quantising the
Hamiltonian (\ref{hamil-01}) suggest that we construct the operator:
\begin{eqnarray}\label{hamil-2}
&&\hat H =-\frac{1}{2}\frac{1}{\sqrt{g}}\partial_a(\sqrt{g}g^{ab}\partial_b)+{\rm tr}V(\lambda)=-\frac{1}{2\Delta}\sum_i{\partial_{\lambda_i}}(\Delta\,\partial_{\lambda_i})-\sum_{i<j}\frac{{\cal L}_{\theta_{ij}}{\cal L}_{\bar\theta_{ij}}}{(\lambda_i-\lambda_j)^2}+{\rm tr}V(\lambda)\ ,\nonumber\\
&&\text{where   }\Delta=\prod_{i>j}(\lambda_i-\lambda_j)^2\ .
\end{eqnarray}
Note that the differential operators ${\cal L}_{\theta_{ij}}$ and
${\cal L}_{\bar\theta_{ij}}$ are left invariant vector fields dual to
the left invariant Maurer-Cartan forms.\footnote{Note also that in
  general there is the freedom to add to the potential a term
  proportional to the scalar curvature, which for one matrix vanishes,
  because the metric (\ref{metric-2}) is flat.} The eigenvalues of the
operator (\ref{hamil-2}) determine the energy of the system. Therefore
we need to solve the corresponding Schr\"{o}dinger equation:
\begin{equation}
\hat H \Psi=N^2E^{(1)}\Psi\ ,
\end{equation}
where we used the notations of ref. \cite{Brezin:1977sv}.  Given that
the potential depends only on the eigenvalues of the matrix we
consider separate angular and eigenvalue dependence. Furthermore our goal is only the 
ground state energy which will be given by the angular independent 
wave function $\Psi(\lambda_1\dots\lambda_N)$, which is
function only of the eigenvalues. Following
ref.~\cite{Brezin:1977sv} we consider the redefinition:
\begin{equation}\label{new-wave-1}
\phi(\lambda_1\dots\lambda_N)=\prod_{i>j} (\lambda_i-\lambda_j)\Psi(\lambda_1\dots\lambda_N)\ .
\end{equation}
Note that in order for the original wave function
$\Psi(\lambda_1\dots\lambda_N)$ to be analytic as
$\lambda_i\to\lambda_j$, we need the new wave function
$\phi(\lambda_1\dots\lambda_N)$ to be completely antisymmetric,
therefore the quantisation of the model is equivalent to the
quantisation of $N$  fermions  in the central
potential $V$. Indeed, the Schr\"{o}dinger equation for the wave
function (\ref{new-wave-1}) becomes \cite{Brezin:1977sv}:
\begin{equation}
\sum_i\left(-\frac{1}{2}\frac{\partial^2}{\partial\lambda_i^2}+V(\lambda_i)\right)\phi=N^2\,E^{(1)}\phi\ .
\end{equation} 
The antisymmetric wave function $\phi$ can be constructed as the Slater determinant:
\begin{equation}
\phi(\lambda_1,\dots,\lambda_N)=\frac{1}{\sqrt{N!}}\sum_{i_1,\dots,i_N}\varepsilon_{i_1\dots i_N}f_{i_1}(\lambda_1)\dots f_{i_N}(\lambda_N)\ ,
\end{equation}
where $f_i$ are individual wave functions normalised to one and satisfying:
\begin{equation}
\left[-\frac{1}{2}\frac{\partial^2}{\partial\lambda^2}+V(\lambda)\right]f_i(\lambda)=e_i f_i(\lambda)
\end{equation}
and $\sum_i e_i=N^2E^{(1)}$, since we are interested in the ground
state of the system we choose the individual energies $e_i$ to label
the first $N$ excited states of the individual hamiltonian.  Now the
distribution of the eigenvalues can be obtained from the modulus of
the wave function $\phi$ by integrating over all but one of the
eigenvalues:
\begin{eqnarray}
\rho(\lambda)&=&\int\prod_{k=2}^{N}d\lambda_k\, |\phi(\lambda,\lambda_2,\dots\lambda_N)|^2=
\sum_{i_2,\dots,i_N}\varepsilon_{i,i_2\dots,i_N}\varepsilon_{ji_2\,\dotsi_N}\frac{f_i(\lambda)f^*_j(\lambda)}{N!}  \nonumber\\
&=&\frac{1}{N}\sum_i|f_i(\lambda)|^2\ .
\end{eqnarray}
Therefore, we see that the distribution of the eigenvalues is given as
an average over the probability distributions determined by the
individual wave functions $f_i$. The key idea to obtain a closed form
expression for the distribution $\rho$ valid at large $N$, is to
notice that at large $N$ the semi-classical approximation to the
individual wave function is well justified and in the limit
$N\to\infty$ becomes exact.  Indeed, rescaling the eigenvalues, and
redefining the wave functions, the potential and the individual
energies via:
\begin{equation}
\lambda_i\to\sqrt{N}\,\lambda_i,~~~\tilde f_i(\lambda)={N^{1/4}}\,f_i(\sqrt{N}\lambda)\,,~~~ u(\lambda) =\frac{1}{N}\,V(\sqrt{N}\,\lambda)\ ,~~~\varepsilon_i =\frac{e_i}{N}
\end{equation}
we arrive at:
\begin{equation}
\left[-\frac{1}{2N^2}\frac{\partial^2}{\partial\lambda^2}+u(\lambda)\right]\tilde f_i(\lambda)=\varepsilon_i \,\tilde f_i(\lambda)\ .
\end{equation}
We see that we have an effective Planck constant $\hbar\propto 1/N$,
and for large $N$ we can use a WKB approximation for the wave
functions\cite{LandauLifshitzQM:1965} $\tilde f_i(\lambda)$:
\begin{eqnarray}
\tilde f_i(\lambda)&=&\frac{C_1}{\sqrt{p(\lambda)}}\cos\left(N\,\int\limits_a^\lambda d\lambda'\,p(\lambda')-\frac{\pi}{4}\right)\ ,\\
p(\lambda)&=&\sqrt{2(\varepsilon_i-u(\lambda))}
\end{eqnarray}
for $a\leq\lambda\leq b$, where $a$ and $b$ are the two turning points
of the classical trajectory. The normalisation coefficient $C_1$ is
fixed by the condition:
\begin{equation}
\int\limits_a^b d\lambda |\tilde f_i(\lambda)|^2\approx \frac{C_1^2}{2}\int\limits_a^b\frac{d\lambda}{p(\lambda)} =\frac{\pi\,C_1^2}{2\,\omega_i}=1\ ,
\label{normalisation}
\end{equation}
where $\omega_i=2\pi/T_i$ is the frequency of the classical motion
depending on the energy $\varepsilon_i$, and we have used that  
for large $N$ to leading order we have 
$\cos^2(N \int^\lambda d\lambda' p(\lambda'))\approx1/2$ under the integral in (\ref{normalisation}). With this
normalisation, and approximating again the fast oscillating cosine function with one half, for the absolute value square of the wave function
$\tilde f_i$ at large $N$ we obtain:
\begin{equation}
|\tilde f_i(\lambda)|^2 =\frac{\omega_i}{\pi\sqrt{2(\varepsilon_i-u(\lambda))}}\ .
\end{equation}
Redefining the distribution for the rescaled eigenvalues (so that it
is normalised to one) via:
\begin{equation}
\tilde\rho(\lambda)=\sqrt{N}\,\rho(\sqrt{N}\lambda) =\frac{1}{N}\sum_i|\tilde f_i(\lambda)|^2\ ,
\end{equation}
 we obtain:
\begin{equation}\label{dist-2}
\tilde\rho(\lambda)=\frac{1}{N}\sum_n\,\frac{\omega_n}{\pi\sqrt{2(\varepsilon_n-u(\lambda))}}\ .
\end{equation}
Our final step is to use the Born-Sommerfeld quantisation (for
$\hbar=1/N$):
\begin{equation}
\frac{N}{2\pi}\oint p \,d\lambda =n+\frac{1}{2}
\end{equation}
to obtain:
\begin{equation}
\frac{N}{2\pi}\oint \Delta p_n \,d\lambda =\frac{N \Delta \varepsilon_n }{2\pi}\oint\limits \frac{\partial p_n}{\partial\varepsilon_n} \,d\lambda 
=\frac{N \Delta \varepsilon_n }{2\pi}\oint \frac{d\lambda }{p_n} =\frac{N \Delta \varepsilon_n }{\omega_n} =\Delta n\ .
\end{equation}
The last equality allows us (in the large $N$ limit) to express the
sum over $n$ in (\ref{dist-2}) as a definite integral over
$\varepsilon$:
\begin{equation}
\tilde\rho(\lambda)=\sum_n\,\frac{\Delta\varepsilon_n}{\pi\sqrt{2(\varepsilon_n-u(\lambda))}}=\frac{1}{\pi}\int\limits_{u(\lambda)}^{\epsilon_f}\frac{d\varepsilon}{\sqrt{2(\varepsilon-u(\lambda))}}=\frac{1}{\pi}{\sqrt{2(\varepsilon_f-u(\lambda))}}\ ,
\end{equation}
which is our final expression for the distribution $\tilde\rho$ and
agrees with the result in ref. \cite{Brezin:1977sv}, as it should,
since we have only made the semiclassical analysis more explicit.

\subsection{Gaussian potential}

Let us now focus on the specific case of a gaussian potential
$V(X)=X^2/2$. In this case the rescaled individual wave functions
$\tilde f_n$ satisfy the Schr\"{o}dinger equation for the one
dimensional harmonic oscillator:
\begin{equation}
\left[-\frac{1}{2N^2}\frac{\partial^2}{\partial\lambda^2}+\frac{1}{2}\lambda^2\right]\tilde f_n(\lambda)=\varepsilon_n \,\tilde f_n(\lambda)\ ,
\end{equation}
where:
\begin{equation}
\tilde f_n(\lambda) =\frac{H_n(\lambda)e^{-N\,\lambda^2/2}}{\sqrt{\pi^{1/2}2^n n!}};~~~\varepsilon_n=\frac{(1/2+n)}{N};
\end{equation}
and $H_n(\lambda)$ are the Hermite polynomials. The corresponding
multi-particle wave function is given by:
\begin{eqnarray}
\tilde \phi(\lambda_1\dots\lambda_N)&=&C_N\sum_{i_1,\dots,i_N}H_{i_1}(\lambda_1)\dots H_{i_N}(\lambda_N)\, e^{-N\,\sum_i {\lambda_i^2}/{2}}\nonumber\\
&=&{2^{\frac{N(N-1)}{2}}C_N}\prod_{i>j}(\lambda_i-\lambda_j)\,e^{-N\,\sum_i {\lambda_i^2}/{2}}\ ,\label{eqn-harm-ferm}
\end{eqnarray}
where $C_N^{-2}=N!\prod\limits_{n=0}^{N-1}(\pi^{1/2}2^n n!)$ and we
have used the properties of the Hermite polynomials to arrive at the
last equality. Equation (\ref{eqn-harm-ferm}) suggests that the bosonic
wave function $\tilde\psi$ before the change of variables
(\ref{new-wave-1}) is given by:
\begin{equation}
\tilde\Psi(\lambda_1\dots\lambda_N)={2^{\frac{N(N-1)}{2}}C_N}\,\prod_{i}e^{-N\,{\lambda_i^2}/{2}}\ .
\end{equation}
Therefore, the bosonic wave function of the ground state of the system
is a product of individual wave functions.  We could have
easily guessed this result even directly in the bosonic description
without introducing the new wave function (\ref{new-wave-1}). Note
that this is true only for the case of a gaussian potential. In fact,
in the next section we will show that this holds even for the case of
more than one commuting matrices with a gaussian potential~\cite{Berenstein:2005aa}.
\section{Commuting matrix model}

In this section we focus on the quantisation of the commuting matrix model. While we will set up the problem for general interaction potential the main findings that we present are for the special case of gaussian potential, first discussed in ref.~\cite{Berenstein:2005aa}. Our starting point is the lagrangian:
\begin{eqnarray}\label{Lag-com}
{\cal L}={\rm tr}\left(\frac{1}{2}\dot {\vec X}^{2}-V(\vec X)\right)\ ,
\end{eqnarray}
where $\vec X$ represent a set of $p$ commuting $N\times N$ hermitian matrices:
\begin{equation}
\vec X = \{X^1,\dots X^p\}~~~{\rm and}~~~[X^{\mu},X^{\nu}]=0~~~{\rm for}~\mu,\nu=0,\dots,p\ .
\end{equation}
Unlike the one matrix model discussed in the previous section we cannot directly write down the corresponding hamiltonian. The problem is that the commuting matrices are not independent and the system is constrained. This is why we will perform the quantisation in ``curved'' coordinates by using the parameterisation:
\begin{equation}
X^{\mu} =U \Lambda^{\mu} U^{\dagger}\ ,~~~{\rm for}~\mu=1,\dots,p\ ,
\end{equation}
where $\Lambda^{\mu}$ are real diagonal matrices consisting of the eigenvalues of the commuting matrices $X^{\mu}$ and $U$ is an unitary matrix. Note that $U$ is defined modulo a right multiplication by a diagonal unitary matrix. Taking the quotient with respect to this equivalence leaves only $N^2-N$ independent real components of the unitary matrix $U$, which together with $p N$ real degrees of the eigenvalue matrices $\Lambda^{\mu}$ suggests that the space of the  $p$ commuting matrices is an $N^2+(p-1)N$ dimensional subspace in the $p N^2$ space of independent hermitian matrices.  Using the natural flat metric in the space of independent hermitian matrices, for the induced metric on the space of commuting matrices we obtain:
\begin{equation}
ds^2={\rm tr}\left(d\vec X .d\vec X\right)=\sum_i d\vec\lambda_i+2\sum_{i>j}(\vec\lambda_i-\vec\lambda_j)^2\theta_{ij}\bar\theta_{ij}\ ,\label{metric-com}
\end{equation}
where $\theta =U^{\dagger}dU$ are the Maurer--Cartan left invariant forms and $\vec\lambda_i=\vec\Lambda_{ii}$ are the eigenvalues of the commuting matrices. In these coordinates the Lagrangian (\ref{Lag-com}) is given by:
\begin{equation}
{\cal L} =\frac{1}{2}{\rm tr}\left(\dot{\vec\Lambda}^2+[\vec\Lambda,\dot\theta]^2    \right)-{\rm tr}V(\vec\Lambda)=\frac{1}{2}\sum_i\dot{\vec\lambda}_i^2+\sum_{i<j}(\vec\lambda_i-\vec\lambda_j)^2\dot\theta_{ij}\bar{\dot\theta}_{ij}-\sum_iV(\vec\lambda_i)\ ,
\end{equation}
where $\dot\theta=U^{\dagger}\dot U$. Defining the conjugate momenta:
\begin{equation}
\Pi_{\vec\lambda_i} =\frac{\partial {\cal L}}{\partial \dot{\vec\lambda}_i}=\dot{\vec\lambda}_i\ ,~~\Pi_{\theta_{ij}} =\frac{\partial {\cal L}}{\partial\, \dot{\bar{\theta}}_{ij}}=(\lambda_i-\lambda_j)^2\dot\theta_{ij}\ ,~~\Pi_{\bar{\theta}_{ij}} =\frac{\partial {\cal L}}{\partial\, \dot{\theta}_{ij}}=\bar{\Pi}_{\theta_{ij}}\ , \text{for }i<j\ .
\end{equation}
For the corresponding classical Hamiltonian we obtain: 
\begin{equation}\label{hamil-1}
H=\frac{1}{2}\sum_i \Pi_{\lambda_i}^2+\sum_{i<j}\frac{\Pi_{\theta_{ij}}\Pi_{\bar{\theta}_{ij}}}{(\lambda_i-\lambda_j)^2}+\sum_iV(\lambda_i)=\frac{1}{2}g^{ab}\Pi_a\Pi_b+\sum_iV(\lambda_i)\ ,
\end{equation}
where $a,b$ are a collective indices for the $\lambda_i^{\mu},
\theta_{ij}$ and $\bar\theta_{ij}$ components of the inverse of the
metric (\ref{metric-com}). So far the analysis was a simple
generalisation of the one performed in the one matrix model
case. However, note that the unlike the metric (\ref{metric-2}), which
is just a flat metric written in curved coordinates, the metric
(\ref{metric-com}) is the induced metric on the subspace of commuting
matrices and in general there is no reason to expect that it is flat
and as it turns out it isn't. It is a well known problem in quantum
mechanics in a curved space that there is an ambiguity in the definition
of the quantum Hamiltonian. In particular there is the freedom to add
a term proportional to the scalar curvature to the potential. The
coefficient in front of this term is not determined, however if one
demands a path integral quantisation with the naive classical
Hamiltonian, this coefficient is fixed to $1/6$. Alternatively one
could introduce an appropriate such term in the Lagrangian so that the
term generated by the measure of the path integral cancels it
\cite{DeWitt:1957at,Cheng:1972yb} and one ends up with a coefficient
zero in the quantum Hamiltonian. It is the quantum system that is more
fundamental and since such a term is induced by the path integral
measure then, when the system is an effective low energy Lagrangian,
it seems natural to allow the coefficient to take on any
value. Our strategy will be to keep this coefficient general and
explore the implications of varying it. Thus we consider the
Hamiltonian.
\begin{eqnarray}\label{hamil-com}
&&\hat H =-\frac{1}{2}\frac{1}{\sqrt{g}}\partial_a(\sqrt{g}g^{ab}\partial_b)+{\rm tr}V(\vec\Lambda)+\frac{\xi}{6}{\cal R}\nonumber \\
&&=-\frac{1}{2\Delta}\sum_i{\partial_{\vec\lambda_i}}.(\Delta\,\partial_{\vec\lambda_i})-\sum_{i<j}\frac{\partial_{\theta_{ij}}\partial_{\bar\theta_{ij}}}{(\vec\lambda_i-\vec\lambda_j)^2}+{\rm tr}V(\vec\Lambda)+\frac{\xi}{6}{\cal R}\ ,
\end{eqnarray}
where $\Delta=\prod_{i>j}(\vec\lambda_i-\vec\lambda_j)^2$ and ${\cal R}$ is the Ricci scalar curvature of the metric (\ref{metric-com}) for which one we obtain the expression:
\begin{eqnarray}\label{curv}
{\cal R}&=&-4(p-1)\sum_{i\neq j}\frac{1}{(\vec\lambda_i-\vec\lambda_j)^2}-3\sum_{i\neq j\neq k}\frac{(\vec\lambda_i-\vec\lambda_j)}{(\vec\lambda_i-\vec\lambda_j)^2}.\frac{(\vec\lambda_i-\vec\lambda_k)}{(\vec\lambda_i-\vec\lambda_k)^2}=\nonumber\\
&=&-(4p+3N-10)\sum_{i\neq j}\frac{1}{(\vec\lambda_i-\vec\lambda_j)^2}+\frac{3}{2}\sum_{i\neq j\neq k}\frac{(\vec\lambda_j-\vec\lambda_k)^2}{(\vec\lambda_i-\vec\lambda_j)^2(\vec\lambda_i-\vec\lambda_k)^2}\ .
\end{eqnarray}
As one can see the curvature can be written with negative two body and positive three body interactions, the net effect in the large $N$ limit however is always a negative curvature. A fact which will prove crucial for the quantisation of the model. 

Our next step is to consider a wave function $\Psi$, which is
invariant under $SU(N)$ gauge transformation and thus independent on
the $\theta$'s. The resulting Schr\"{o}dinger equation has a term
involving a first derivative of the wave function, which we can
discard by considering the higher dimensional analogue of the change
of variables (\ref{new-wave-1}):
\begin{equation}\label{change-com}
\Phi(\vec\lambda_1,\dots,\vec\lambda_N)=\prod_{i>j}|\vec\lambda_i-\vec\lambda_j|\,\Psi(\vec\lambda_1,\dots,\vec\lambda_N)\ .
\end{equation}
Taking into account equations (\ref{hamil-com}), (\ref{curv}) and (\ref{change-com}) for the Schr\"{o}dinger equation satisfyed by $\Phi$ we obtain:
\begin{eqnarray}\label{Shrod-phi}
&&-\frac{1}{2}\sum_i\frac{\partial^2\Phi}{\partial\vec\lambda_i^2}+\frac{3-4\xi}{6}(p-1)\sum_{i\neq j}\frac{\Phi}{(\vec\lambda_i-\vec\lambda_j)^2}+\frac{1-\xi}{2}\sum_{i\neq j\neq k}\frac{(\vec\lambda_i-\vec\lambda_j)}{(\vec\lambda_i-\vec\lambda_j)^2}.\frac{(\vec\lambda_i-\vec\lambda_k)}{(\vec\lambda_i-\vec\lambda_k)^2}\Phi+\nonumber\\
&&+\sum_i V(\vec\lambda_i)\Phi =N^2\,E\,\Phi\ .
\end{eqnarray}
Few comments are in order. Notice that the change of variables (\ref{change-com}) does not affect the symmetry of the wave function and the new wave function $\Phi$ is symmetric thus describing bosons.\footnote{Note that for $p=2$ an alternative quantisation is
also possible. One could take the analytic square root of the eigenvalue difference by going
to complex coordinates and make contact with anyonic systems~\cite{Wilczek:1982wy}.} This is in contrast to the one dimensional case, where the Vandermonde determinant in (\ref{new-wave-1}) implied that the new wave function is antisymmetric. Therefore, the problem of quantising the commuting model is equivalent to finding a bosonic wave function satisfying the Schr\"{o}dinger equation (\ref{Shrod-phi}), which describes $p$-dimensional bosonic particles subjected to the central potential $V$ and with two and three body interaction terms. 

Note that for $\xi \geq 1$ the two and three body interactions are attractive, which naively suggests that the model will be unstable. One can of course imagine that the kinetic energy of the particles may stabilise the system. However, as we are going to demonstrate, doing so would require that the number of commuting matrices go to infinity in the large $N$ limit. Indeed, the instability arises, when the eigenvalues are close to each other. In this limit the contribution from the two and tree-body interactions dominate and we can ignore the central potential. Let us first consider the case $\xi$ strictly grater than one, $\xi >1$. Assuming a fixed typical scale for the spread of the eigenvalues $|\vec\lambda_i-\vec\lambda_j|\approx \Delta R$ and using equations (\ref{Shrod-phi}) and (\ref{curv}), for the leading order contribution to the potential energy (at large $N$) we obtain: 
\begin{equation}\label{pot-en}
E_{\rm pot}=-\frac{\xi-1}{2\Delta R^2} N^3 +O(N^2)\ .
\end{equation}
On the other side, in $p$ dimensions the typical scale along one of the coordinates $\Delta x$ is related to $\Delta R$ via: $\Delta x^2 =\Delta R^2/p$. This suggests that the uncertainty of the momentum along one of the coordinates is $\Delta q \propto \hbar /\Delta x$ and therefore the kinetic energy is of order:
\begin{equation}\label{kin-en}
E_{\,\rm kin}=N\,p\, \Delta q^2\propto N\,p^2/\Delta R^2 \ .
 \end{equation}
Stabilising the model would require $E_{\rm kin} \sim | E_{\rm pot}|$, comparing equations (\ref{pot-en}) and (\ref{kin-en}) we conclude that this would imply $p\sim N$, and hence in the large $N$ limit the number of matrices should grow proportional to the size of the matrices. Clearly, for models with fixed number of commuting matrices $p$, the kinetic energy is not sufficient to stabilise the system for arbitrary large $N$ and the large $N$ limit of these models does not have a ground state for $\xi >1$.

 Similar result can be obtained for the case $\xi=1$, when the three body interaction term vanishes. In this case the potential energy is dominated by the two body potential and to leading order in $N$ one obtains:
 \begin{equation}
E_{\rm pot} =- (p-1)N^2/\Delta R^2\ .
\end{equation} 
 Comparing this expression for the potential energy with the expression for the kinetic energy (\ref{kin-en}), which  is still valid, one again arrives at the result that stabilisation of the model in the large $N$ limit requires $p\sim N$ and the commuting matrix model with fixed number of commuting matrices is unstable for $\xi=1$. 
 
Let us demonstrate how this instability occurs in the special case of attractive gaussian potential.
\subsection{Gaussian potential}
In this subsection we analyse the case of gaussian potential
$V(\lambda)=\frac{1}{2}\lambda^2$, first studied in
ref.~\cite{Berenstein:2005aa} for the case $\xi=0$. For $\xi=0$ one
can show that the ground state wave function is given by
\cite{Berenstein:2005aa}:
\begin{equation}
\Phi(\vec\lambda_1,\dots,\vec\lambda_N)=\prod_{i>j}|\vec\lambda_i-\vec\lambda_j|\,e^{-\frac{1}{2}\sum_i\vec\lambda_i^2}\ ,
\end{equation}
which is the naive generalisation to higher dimensions of the wave
function (\ref{eqn-harm-ferm}). Note that this is an exact result at
any $N$. For general $\xi$ it seems that we don't have analytic access
to the exact result, however we can still extract the ground state
wave function and the corresponding eigenvalue distribution in the
large $N$ limit. To see this let us rescale the eigenvalues by:
$\vec\lambda_i\to N^{1/2}\vec\lambda_i\,$. After dividing equation
(\ref{Shrod-phi}) by $N$ we obtain:
\begin{eqnarray}\nonumber
&\frac{1}{N}\sum_i&\left\{ -\frac{1}{2N^2}\frac{\partial^2\tilde\Phi}{\partial\vec\lambda_i^2}+\frac{(3-4\xi)(p-1)}{6 N^2}\sum_{\jj\limits_{j\neq i}}\frac{\tilde\Phi}{(\vec\lambda_i-\vec\lambda_j)^2} +\frac{1-\xi}{2N^2}\sum_{j\neq k}\frac{(\vec\lambda_i-\vec\lambda_j)}{(\vec\lambda_i-\vec\lambda_j)^2}.\frac{(\vec\lambda_i-\vec\lambda_k)}{(\vec\lambda_i-\vec\lambda_k)^2}\tilde\Phi+ \right. \\
&&\left. +\frac{1}{2}\vec\lambda_i^2\tilde\Phi    \right\}=E\,\tilde\Phi\ . \label{shord-resc-phi}
\end{eqnarray}
Note that the two body interaction (the second) term in equation (\ref{shord-resc-phi}) is of order $1/N$ relative to the other potential terms. Therefore, at least naively it should not contribute to the leading order behaviour of the wave function in the large $N$ limit. With this in mind one can show that the following ansatz for the wave function:
\begin{equation}\label{lead-wave}
\tilde\Phi(\vec\lambda_1,\dots,\vec\lambda_N)=\prod_{i>j}|\vec\lambda_i-\vec\lambda_j|^{\sqrt{1-\xi}}\,e^{-\frac{N}{2}\sum_i\vec\lambda_i^2}\ ,
\end{equation}
satisfies equation (\ref{shord-resc-phi}) to leading order in the large $N$ expansion and for the ``stabilised'' energy $E$ we obtain:
\begin{equation}
E=\frac{1}{2}\sqrt{1-\xi}+O(1/N)\ .
\end{equation}
To verify the consistency of this approximation we treat the two body
interaction term as a perturbation $\delta V$ to the potential of
relative order $1/N$, therefore the corresponding correction to the
energy $\delta E $ would also be of order $1/N$ and we conclude that
as long as $E\sim 1$ we can trust the wave function
(\ref{lead-wave}). Note that this suggests that the parameter $\xi$ is
restricted to $\xi<1$, which is consistent with the general arguments
(presented in the previous section) that the model is unstable for
$\xi \geq 1$. Let us comment on the properties of the eigenvalue
distribution for $\xi <1$.

The probability density corresponding to the wave function (\ref{lead-wave}) can be written as:
\begin{equation}
|\tilde\Phi|^2 =\exp\left[-N\sum_i\vec\lambda_i^2+\frac{\sqrt{1-\xi}}{2}\sum_{i\neq j}\log(\vec\lambda_i-\vec\lambda_j)^2\right]\ .
\end{equation}
The corresponding eigenvalue distribution:
\begin{equation}
\rho(\vec\lambda)=\int\prod_{i=2}^N d^p\lambda_i\,|\tilde\Phi(\vec\lambda,\vec\lambda_2,\dots,\vec\lambda_N)|^2
\end{equation}
is then equivalent to the eigenvalue distribution of the commuting matrix model described by the action:
\begin{equation}
S=N\sum_i\vec\lambda_i^2 -\frac{\sqrt{1-\xi}}{2}\sum_{i\neq j}\log(\vec\lambda_i-\vec\lambda_j)^2\ ,
\end{equation}
which modulo normalisation constants was analysed in refs.~\cite{Berenstein:2005jq}-\cite{Filev:2014jxa}. The corresponding saddle point equation is given by:
\begin{equation}
\frac{1}{\sqrt{1-\xi}}\,\vec\lambda_i =\frac{1}{N}\sum_{\jj\limits_{j\neq i}}\frac{\vec\lambda_i-\vec\lambda_j}{(\vec\lambda_i-\vec\lambda_j)^2}\ ,
\end{equation}
comparing this to equation (2.6) in ref.~\cite{Filev:2014jxa} we
conclude that all of the results of \cite{Filev:2014jxa} for gaussian
potential are valid in here provided we multiply the radius of the
distribution by a factor of $(\frac{1-\xi}{4})^{1/4}$. Therefore, we obtain that
for $2\leq p\leq 4$ the radius of the distribution is given by
\cite{Filev:2014jxa}:
\begin{equation}
R_p^2=\frac{2}{p}\sqrt{1-\xi}\ ,
\end{equation}
being a disk for $p=2$, of the form $\sim (R_3^2-\lambda^2)^{-1/2}$ for $p=3$, and for $p\geq 4$ a spherical shell of radius $R_p=(\frac{1-\xi}{4})^{1/4}$.

Note that at $\xi=1$ the radius of the distribution collapses to
zero. This suggests that the model is unstable in this regime, which
is consistent with the analysis of the previous section.

There is another interesting set of values of the parameter $\xi$ for
which the ground state wave function (\ref{lead-wave}) is exact for
any $N$ \cite{Khare:1996jn,Khare:1997yz}. Indeed, the order $1/N$ term
that we ignored when substituting equation (\ref{lead-wave}) into
equation (\ref{shord-resc-phi}) is given by:
\begin{equation}
\frac{(6-3p)(\sqrt{1-\xi}-1)+(7-4p)\xi}{6N}\frac{1}{N^2}\sum_{i\neq j}\frac{1}{(\vec\lambda_i-\vec\lambda_j)^2}\ .
\end{equation}
Imposing that the coefficient in front of the double sum vanishes we obtain the following set of values of the parameter $\xi$:
\begin{equation}
\xi = 0\ ,~~~{\rm or}~~~\xi=\frac{3(5p-8)(p-2)}{(4p-7)^2}\ ,
\end{equation}
for which the ground wave function is exact for any $N$. The case
$\xi=0$ was analysed in ref.~\cite{Berenstein:2005aa}, while for the
other case we have to check that it is consistent with the stability
condition $\xi<1$, obtained above. Indeed on can check that for $p\geq
2$ the second solution is in the range $0\leq \xi <\frac{15}{16}$,
which is consistent.

\section{Conclusions}
We have studied a quantum mechanical system comprised of a set of $p$
commuting matrices whose Hamiltonian has $SO(p)$ symmetry. Our starting 
point was the curved configuration space for commuting matrices. 
We found that this space is flat only for the one matrix, with in general
a scalar curvature  (\ref{curv-app}) which is singular and 
negative for all $p>1$ when pairs of eigenvalues approach one another.

As a byproduct we caculated the scalar curvature, of a generic flag
manifold, where the different radii squashing the standard flag,
$\Delta_{ij}$, are given by the rotationally invariant differences of
eigenvalues $\Delta^2_{ij}=(\vec{\lambda}_i-\vec{\lambda}_j)^2$ . The
flag curvature is given by the relatively simple expression
$${\cal R}_x=N\,\sum_{i\neq j}\frac{1}{\Delta_{ij}^2}-\frac{1}{2}\sum_{i\neq j\neq k} \frac{\Delta_{jk}^2}{\Delta_{ij}^2\,\Delta_{ik}^2}\ ,$$
see (\ref{flag-curv}).

The negative curvature comes from the extrinsic curvature
contribution, ${\cal K}_\lambda$, to the total curvature and is given
in (\ref{flag-curv}).

We, unfortunately, were unable to solve for the ground state
wavefunction for a generic potential. However, we established that
with the curvature coefficient $\xi\ge 1$ the system has no finite
energy ground state and so, irrespective of the potential the
eigenvalues collapse.  When the coupling $\xi$ is less than one
we treated the case of a quadratic potential and found the ground
state energy $E=\frac{1}{2}\sqrt{1-\xi}+O(1/N)$ and the wavefunction
was Gaussian with an eigenvalue repulsion coefficient given by
$\Delta_{ij}^{\sqrt{1-\xi}}$.

One advantage of considering generic $\xi$ is that it allows us to
make contact with the literature for higher dimensional
Calogero-Sutherland-Marchioro type models, since the Hamiltonian for
our system (\ref{shord-resc-phi}) with quadratic potential is of this
type (see equations (7) of \cite{Khare:1996jn} and (3) of
\cite{Khare:1997yz}).

In a fashion analogous to $p=1$, where both Fermionic and Bosonic
quantisations were possible, for $p=2$ an alternative quantisation is
also possible.  Though we have not pursued this line of investigation,
in the case of $p=2$ it is interesting to note that one could have
taken the analytic square root of the eigenvalue difference by going
to complex coordinates and make contact with anyonic systems
\cite{Wilczek:1982wy}. The Hamiltonian in this case can be mapped into
a multi-anyon system as in eqn (3) of \cite{Date:1992uu}.

The novel r\^ole played by $p=4$ is that once the number of commuting
matrices is four or more the eigenvalues form infinitesimally thin
shells. We establish this when the potential is Gaussian but we
suspect that it may be a more general result. The study of this
question warrants another study and we hope to return to this question
in the future.

\appendix 

\section{Scalar curvature of the space of commuting hermitian matrices}\label{Appendix A}
In this appendix we derive the expression for the scalar curvature of the space of commuting hermitian matrices (\ref{curv}). Let us consider the induced metric on this space given in equation (\ref{metric-com}), which we duplicate bellow:
\begin{equation}
ds^2=\sum_i d\vec\lambda_i+2\sum_{i>j}(\vec\lambda_i-\vec\lambda_j)^2\theta_{ij}\bar\theta_{ij}\ .\label{metric-com-app}
\end{equation}
A key observation is that the dependence of the metric on the directions of the flag manifold is entirely in terms of the left invariant Maurer--Cartan forms $\theta_{ij}$ (more precisely the off diagonal ones for $i\neq j$). This implies that the homogeneous structure of the flag is preserved and at fixed eigenvalues (fixed $\vec\lambda_i$) all points on the flag are equivalent in the sense that the vicinity of all points looks the same. This implies that the sacral curvature should not depend on the flag directions and can be calculated locally at any point of the flag. The most convenient choice is to calculate the scalar curvature at the origin of the flag manifold. Since the curvature involves up to second derivatives of the metric we conclude that it is sufficient to know the explicit parametrisation of the metric up to second order near the origin of the flag. 

Let us describe our choice of parametrisation. The parametrisation of a general element of $U(N)$ near the unity is given by $U=e^{i x}$, where $x$ is an $N\times N$ hermitian matrix with $N^2$ independent degrees of freedom. The flag manifold is obtained by factoring with respect to all diagonal unitary matrices and hence is described by $N(N-1)$ independent parameters $\omega^a$, The flag can then be embedded in the $U(N)$ manifold by the parametrisation $x(\omega)$. In general the closed form of such a parametrisation is rather complex, however it can be constructed easily locally -- in the vicinity of the unity element of $U(N)$. Indeed, to third order in $x$ we have:
\begin{equation}
U=\hat 1 +i\,x-\frac{1}{2}x.x-\frac{i}{6}x.x.x +O(x^4) 
\end{equation}
and using the definition $\theta = U^{\dagger}dU$ for the off diagonal Maurer--Cartan forms we obtain:
\begin{equation}
\theta_{ij}= i\,dx_{ij}-\frac{1}{2}\{x, dx\}_{ij}-\frac{i}{6}\{x^2,dx\}_{ij}-\frac{i}{6}(x.dx.x)_{ij}+O(x^3)\ .
\end{equation}
One can see that at the origin ($x=0$) the off diagonal Muarer--Cartan forms are given by $i \,dx_{ij}$ and are hence parametrised by the off diagonal elements of $x$, exactly $N(N-1)$ real degrees of freedom. Therefore there us always a sufficiently small vicinity of the unit element, where the off diagonal Muarer--Cartan forms can be parametrised by the off diagonal elements of the hermitian matrix $x$ and the embedding of the flag in this coordinates is simply given by $x_{ii}=0$ and $x(\omega)_{ij}=\omega_{ij}$ for $i\neq j$. Our strategy in calculating the scalar curvature will be to treat the the off diagonal elements $x_{ij}$ as complex coordinates. For the components of the metric to second order in $x_{ij}$ we obtain:
\begin{eqnarray}
&&G_{\lambda_i^{\mu},\,\lambda_j^{\nu}}=\delta_{ij}\,\delta_{\mu\nu}\ ,\\
&&G_{x_{ij},\,x_{ml}}=\Delta_{ij}^2\,\delta_{il}\,\delta_{jm} +\frac{i}{2}(x_{li}\,\delta_{jm}-x_{jm}\,\delta_{il})(\Delta_{ij}^2-\Delta_{ml}^2)+\nonumber\\
&&+\frac{1}{12}\left[(3\Delta_{is}^2-2\Delta_{ij}^2-2\Delta_{im}^2)\,x_{js}\,x_{sm}\,\delta_{il}+(3\Delta_{ms}^2-2\Delta_{ml}^2-2\Delta_{im}^2)\,x_{ls}\,x_{si}\,\delta_{mj}\right]+\nonumber\\
&&+\frac{1}{12}(4\Delta_{ij}^2-3\Delta_{jl}^2-3\Delta_{im}^2+4\Delta_{ml}^2)\,x_{li}\,x_{jm}\ , \\
&&G_{\lambda_i^{\mu},\,x_{lm}} =G_{x_{lm},\,\lambda_i^{\mu}}=0\ ,
\end{eqnarray}
where $\Delta_{ij}^2\equiv (\vec\lambda_i-\vec\lambda_j)^2$. The calculation of the components of the associated Levi-Chevita connection is straightforward, the non-zero components (up to first order in $x$) are: 
\begin{eqnarray}
\Gamma_{x_{ks},\,x_{qp}}^{\,\lambda_i^{\mu}}&=&-(\delta_{ik}-\delta_{is})\,\delta_{kp}\,\delta_{qs}\,\Delta_{ks}^{\mu}+\frac{i}{2}(x_{sq}\,\delta_{kp}-x_{pk}\,\delta_{qs})((\delta_{ik}-\delta_{is})\Delta_{ks}^{\mu}+(\delta_{iq}-\delta_{ip})\Delta_{pq}^{\mu})\nonumber \ ,\\
\Gamma_{\lambda_{i}^{\mu},\,x_{qp}}^{\,x_{ks}}&=&(\delta_{iq}-\delta_{ip})\,\delta_{kq}\,\delta_{ps}\,\frac{\Delta_{ks}^{\mu}}{\Delta_{ks}^2}+\frac{i}{2\Delta_{ks}^2}\left(\delta_{ip}\,\delta_{ps}\,x_{kq}\,\Delta_{kq}^{\mu}+\delta_{ik}\,\delta_{kq}\,x_{ps}\,\Delta_{ps}^{\mu}\right)-\nonumber\\
&&-\frac{i}{2\Delta_{ks}^2}(\delta_{ip}-\delta_{is})\,\delta_{kq}\,x_{ps}\,\Delta_{ik}^{\mu}-\frac{i}{2\Delta_{ks}^2}(\delta_{ik}-\delta_{iq})\,\delta_{ps}\,x_{kq}\,\Delta_{ip}^{\mu}\ ,\\
\Gamma_{x_{ks}\,,x_{pq}}^{x_{lm}}&=&\frac{i}{2\,\Delta_{lm}^2}(\delta_{kq}\,\delta_{pl}\,\delta_{sm}-\delta_{lk}\,\delta_{mq}\,\delta_{sp})\left({\Delta_{ks}^2}-{\Delta_{pq}^2}\right)+\frac{1}{12}\,\delta_{kq}\,(x_{sm}\,\delta_{lp}+x_{lp}\,\delta_{ms})+\nonumber\\
&&+\frac{1}{4}\,\delta_{ps}\,(x_{qm}\,\delta_{kl}+x_{lk}\,\delta_{mq})\,\left(\frac{1}{3}+\frac{\Delta_{kp}^2}{\Delta_{km}^2}-\frac{\Delta_{lp}^2}{\Delta_{lq}^2}-\frac{\Delta_{mp}^2}{\Delta_{km}^2}+\frac{\Delta_{pq}^2}{\Delta_{lq}^2}\right)+\nonumber\\
&&+\frac{1}{4}\,\delta_{kq}\,(x_{sm}\,\delta_{lp}-x_{lp}\,\delta_{ms})\,\left(\frac{\Delta_{kl}^2}{\Delta_{lm}^2}-\frac{\Delta_{km}^2}{\Delta_{lm}^2}+\frac{\Delta_{ks}^2}{\Delta_{sp}^2}-\frac{\Delta_{pq}^2}{\Delta_{sp}^2}\right)+\nonumber\\
&&+\frac{1}{4}(x_{sp}\,\delta_{kl}\,\delta_{mq}+x_{qp}\,\delta_{lp}\,\delta_{ms})\left(-\frac{2}{3}+\frac{\Delta_{kp}^2}{\Delta_{lm}^2}-\frac{\Delta_{ks}^2}{\Delta_{lm}^2}-\frac{\Delta_{pq}^2}{\Delta_{lm}^2}+\frac{\Delta_{sq}^2}{\Delta_{lm}^2}\right)\ ,\label{con-flag}
\end{eqnarray}
where $\Delta_{ij}^{\mu}\equiv \lambda_i^{\mu}-\lambda_j^{\mu}$ and again $\Delta_{ij}^2=(\vec\lambda_i-\vec\lambda_j)^2$. 

The next step is to use the standard formula for the Riemman curvatute tensor:
\begin{equation}
{\cal R}^{\rho}_{\sigma\,\mu\nu}=\partial_{\mu}\Gamma^{\rho}_{\nu\sigma}-\partial_{\nu}\Gamma^{\rho}_{\mu\sigma}+\Gamma^{\rho}_{\mu\lambda}\Gamma^{\lambda}_{\nu\sigma}-\Gamma^{\rho}_{\nu\lambda}\Gamma^{\lambda}_{\mu\sigma}\ .
\end{equation}
and the definition for the scalar curvature ${\cal R} =g^{\sigma\nu}\,{\cal R}^{\lambda}_{\sigma\,\lambda\nu}$, to arrive at the expression (\ref{curv}):
\begin{eqnarray}\label{curv-app}
{\cal R}&=&-(4p+3N-10)\sum_{i\neq j}\frac{1}{(\vec\lambda_i-\vec\lambda_j)^2}+\frac{3}{2}\sum_{i\neq j\neq k}\frac{(\vec\lambda_j-\vec\lambda_k)^2}{(\vec\lambda_i-\vec\lambda_j)^2(\vec\lambda_i-\vec\lambda_k)^2}=\nonumber \\
&=&-4(p-1)\sum_{i\neq j}\frac{1}{(\vec\lambda_i-\vec\lambda_j)^2}-3\sum_{i\neq j\neq k}\frac{(\vec\lambda_i-\vec\lambda_j)}{(\vec\lambda_i-\vec\lambda_j)^2}.\frac{(\vec\lambda_i-\vec\lambda_k)}{(\vec\lambda_i-\vec\lambda_k)^2}\ ,
\end{eqnarray}
where we have used the relation:
\begin{equation}
\sum_{i\neq j\neq k}\frac{(\vec\lambda_j-\vec\lambda_k)^2}{(\vec\lambda_i-\vec\lambda_j)^2(\vec\lambda_i-\vec\lambda_k)^2}=2(N-2)\sum_{i\neq j}\frac{1}{(\vec\lambda_i-\vec\lambda_j)^2}-2\sum_{i\neq j\neq k}\frac{(\vec\lambda_i-\vec\lambda_j)}{(\vec\lambda_i-\vec\lambda_j)^2}.\frac{(\vec\lambda_i-\vec\lambda_k)}{(\vec\lambda_i-\vec\lambda_k)^2}\ .
\end{equation}
We can present another useful result. Note that if we keep the eigenvalues fixed, the expression for the connection involving only components along the flag directions, equation (\ref{con-flag}), represents the connection of a flag manifold with constant radii $|\vec\Delta_{ij}|=\sqrt{\Delta_{ij}^2}$. The result in equation (\ref{curv-app}) can be split into contributions from the flag directions ${\cal R}_x$ and contribution from the mixed terms ${\cal K}_\lambda$, the two terms read:
\begin{eqnarray}\label{flag-curv}
{\cal R}_x&=&N\,\sum_{i\neq j}\frac{1}{\Delta_{ij}^2}-\frac{1}{2}\sum_{i\neq j\neq k} \frac{\Delta_{jk}^2}{\Delta_{ij}^2\,\Delta_{ik}^2}\ ,\\
{\cal K}_{\lambda}&=&-(4p-2)\,\sum_{i\neq j}\frac{1}{\Delta_{ij}^2}-4\sum_{i\neq j\neq k}\frac{\vec\Delta_{ij}\,.\,\vec\Delta_{ik}}{\Delta_{ij}^2\,\Delta_{ik}^2}\ .
\end{eqnarray}
Equation (\ref{flag-curv}) is our expression for the curvature of a flag manifold with radii $|\vec\Delta_{ij}|=\sqrt{\Delta_{ij}^2}$.

%

\end{document}